\documentclass[aps,PRB,twocolumn,superscriptaddress,preprintnumbers]{revtex4-2}
\usepackage[T1]{fontenc}
\usepackage{times}
\usepackage{graphicx,color}
\usepackage{amsfonts,amsmath,amssymb,amsbsy}
\usepackage[colorlinks=true, linkcolor=blue, citecolor=blue, urlcolor=blue]{hyperref}
\usepackage{float}

\begin{document}

\title{Detecting the N\'{e}el vector of altermagnet by attaching a
topological insulator\\
and crystalline valley-edge insulator }
\author{Motohiko Ezawa}
\affiliation{Department of Applied Physics, The University of Tokyo, 7-3-1 Hongo, Tokyo
113-8656, Japan}

\begin{abstract}
In order to detect the N\'{e}el vector of an altermagnet, we investigate
topological phases in a bilayer system composed of an altermagnet and a
two-dimensional topological insulator described by the Bernevig-Hughes-Zhang
model. A topological phase transition occurs from a first-order topological
insulator to a trivial insulator at a certain critical altermagnetization if
the N\'{e}el vector of altermagnet is along the $x$ axis or the $y$ axis. It
is intriguing that valley-protected edge states emerge along the N\'{e}el
vector in this trivial insulator, which are as stable as the topological
edge states. We name it a crystalline valley-edge insulator. On the other
hand, the system turns out to be a second-order topological insulator when
the N\'{e}el vector is along the $z$ axis. The tunneling conductance has a
strong dependence on the N\'{e}el vector. In addition, the band gap depends
on the N\'{e}el vector, which is measurable by optical absorption. Hence, it
is possible experimentally to detect the N\'{e}el vector by measuring
tunneling conductance and optical absorption.
\end{abstract}

\date{\today }
\maketitle




Ferromagnets and antiferromagnets are typical examples of magnets. On one
hand, ferromagnets are used for magnetic memory, where the direction of
magnetization stores the nonvolatile memory. The magnetization direction can
be read out electrically\ by using the anomalous Hall effect due to the
time-reversal breaking. However, ferromagnets produce stray field due to the
nonzero net magnetization, which prevents ultra high density integrated
memory. On the other hand, antiferromagnet spintronics\cite%
{Jung,Baltz,Han,Ni,Godin,Kimura,ZhangNeel} have a merit that it is possible
to make ultra high density integrated memory due to the zero net
magnetization. In addition, the flipping speed of antiferromagnets is much
faster than that of ferromagnets\cite{Wu}. However, it is very hard to read
out the N\'{e}el vector because the anomalous Hall effect is absent due to
the combination symmetry of time-reversal symmetry and translational
symmetry. One of the solution is multipolar antiferromagnet\cite{Nakatsuji},
where the anomalous Hall effect is observed.

Recently, altermagnets have attracted rapid growth of interest\cite%
{SmejRev,SmejX,SmejX2}. It is the third-type of magnets in terms of
symmetry. Altermagnets break time-reversal symmetry. Hence, an anomalous
Hall effect emerges\cite{Fak,Tsch,Sato,Leiv}, which can be used for readout
of the $z$ component of altermagnetization. Furthermore, there is no net
magnetization in altermagnets, which may lead to an ultra-high density
integration and high flipping rate of memory. Namely, altermagnets have
merits of both ferromagnets and antiferromagnets. The characteristic feature
of altermagnets is a momentum-dependent band structure for each spin \cite%
{SmejRev,SmejX,SmejX2}. Indeed, momentum dependent band structures are
observed by Angle-Resolved Photo-Emission Spectroscopy (ARPES)\cite%
{Krem,Lee,Fed,Osumi,Lin}. Furthermore, spin current is generated by applying
electric field\cite{Naka,Gonza,NakaB,Bose} owing to the above-mentioned
characteristic band structure. There are only few works on topological
properties\cite{Fer} induced by altermagnets except for Majorana states\cite%
{Zu2023,Li2023,Gho}.

In this paper, analyzing a bilayer system composed of an altermagnet and a
two-dimensional topological insulator, we construct a topological phase
diagram. It has intriguing features. On one hand, a topological phase
transition occurs from a first-order topological insulator to a trivial
insulator at a certain altermagnetization $J_{\text{cr}}$ when the N\'{e}el
vector is along the $x$ axis or the $y$ axis. This trivial insulator is
intriguing since it is characterized by the emergence of edge states
parallel to the N\'{e}el vector, which are as robust as topological edges.
Furthermore, an edge state connects two valleys either in the occupied band
or in the unoccupied band, as is a reminiscence of the valley-protected edge
states in the valley-Chern insulator\cite{Kirch}. Hence, we name it the $x$%
-axis crystalline valley-edge insulator ($x$-CVEI) or the $y$-axis
crystalline valley-edge insulator ($y$-CVEI). On the other hand, the system
becomes a second-order topological insulator (SOTI) when the N\'{e}el vector
is along the $z$ axis. We show that tunneling conductance has a sharp
dependence on the N\'{e}el vector. In addition, the band gap has a
dependence on the N\'{e}el vector, which is measured by optical absorption.
Therefore, the N\'{e}el vector is experimentally determined by combining the
tunneling conductance and the optical absorption spectra.

\textbf{Model:} We analyze the bilayer system where $d$-wave altermagnet is
attached on a two-dimensional topological insulator. The Hamiltonian is
given by%
\begin{equation}
H=H_{\text{BHZ}}+H_{\text{Alter}}.  \label{TotalHamil}
\end{equation}%
The topological insulator is described by the Bernevig-Hughes-Zhang (BHZ)
model\cite{BHZ},%
\begin{equation}
H_{\text{BHZ}}=M\left( k\right) \sigma _{0}\otimes \tau _{z}+\lambda \left(
\sin k_{x}\sigma _{x}\otimes \tau _{x}+\sin k_{y}\sigma _{y}\otimes \tau
_{x}\right) ,  \label{BHZ}
\end{equation}%
with $M\left( \mathbf{k}\right) =m-t\left( \cos k_{x}+\cos k_{y}\right) $,
where $m$ is the mass parameter, $t$ is the hopping parameter, $\lambda $ is
the spin-orbit interaction, $\sigma $ is the Pauli matrix for the spin and $%
\tau $ is the Pauli matrix for the orbital. The BHZ model describes a
topological insulator for $\left\vert m/\left( 2t\right) \right\vert <1$.

The effect of $d$-wave altermagnet is described by the Hamiltonian\cite%
{SmejRev,SmejX,SmejX2,Zu2023,Gho,Li2023} 
\begin{equation}
H_{\text{Alter}}\left( \mathbf{k}\right) =J\left( \cos k_{x}-\cos
k_{y}\right) \left( \mathbf{s}\cdot \mathbf{\sigma }\right) \otimes \tau
_{x},  \label{HJ}
\end{equation}%
with $\mathbf{s}=\left( \sin \theta \cos \phi ,\sin \theta \sin \phi ,\cos
\theta \right) $, where $J$ is the magnitude of altermagnetization and $J%
\mathbf{s\ }$is the N\'{e}el vector. The N\'{e}el vector is controlled by
spin-orbit torque\cite{Bai2023,HanScience} or spin-split torque\cite%
{Karube,Bai2022}. The characteristic feature of altermagnet is that the
magnetization has a momentum dependence as in the Hamiltonian (\ref{HJ}).

\begin{figure}[t]
\centerline{\includegraphics[width=0.48\textwidth]{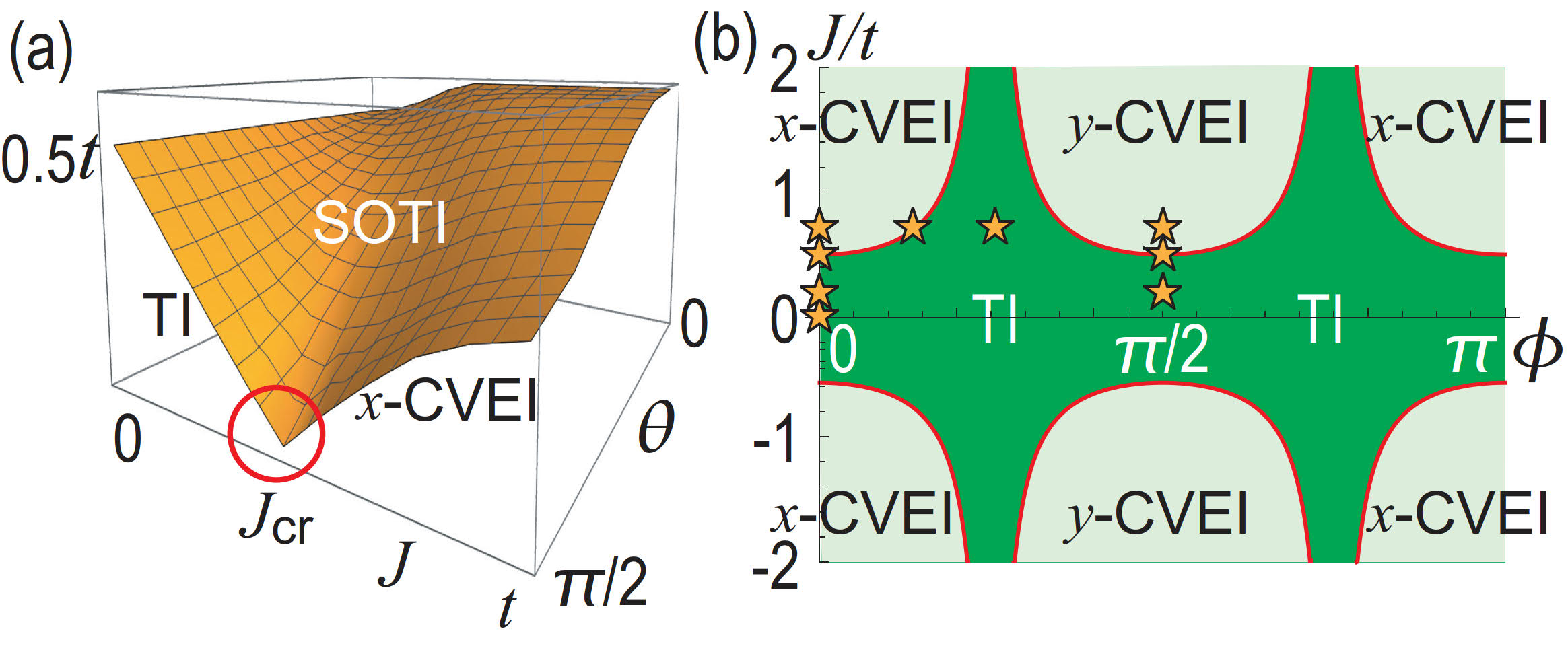}}
\caption{ (a) Bulk band gap in the $J$-$\protect\theta $ plane with $0\leq
J\leq t$ and $\protect\pi /2\geq \protect\theta \geq 0$, where we have set $%
\protect\phi =0$. The band gap closes at $J=J_{\text{cr}}$ and $\protect%
\theta =\protect\pi /2$. (b) Topological phase diagram in the $\protect\phi $%
-$J$ plane with $0\leq \protect\phi \leq \protect\pi $ and $-2\leq J/t\leq 2$%
, where we have set $\protect\theta =\protect\pi /2$. The topological phase
boundaries are analytically given by Eq.(\protect\ref{Jcr2}). A star
represents a point at which the band structure in ribbon geometry is shown
in Fig.\protect\ref{FigRibbon}. We have set $m=t$ and $\protect\lambda =0.5t$%
. }
\label{FigGap}
\end{figure}

\textbf{Band gap:} We study the bulk band gap. If the N\'{e}el vector is
along the $x$ axis, the energy spectrum is given by%
\begin{equation}
E^{2}=M^{2}\left( \mathbf{k}\right) +\left( \lambda \sin k_{x}+J\left( \cos
k_{x}-\cos k_{y}\right) \right) ^{2}+\lambda ^{2}\sin ^{2}k_{y}.
\end{equation}%
The minimum of the band gap is taken at $k_{y}=0$. The solution $M\left(
k_{x},0\right) =0$ is $k_{x}=\pm \arccos \left( m/t-1\right) $. The gap
closing condition is 
\begin{equation}
\left\vert J_{\text{cr}}/\lambda \right\vert =1/\sqrt{2t/m-1}.  \label{Jcr1}
\end{equation}%
We show the band gap in the $J$-$\theta $\ plane at $\phi =0$ in Fig.\ref%
{FigGap}(a). The band gap closes at $J=J_{\text{cr}}$\ when $\theta =\pi /2$.

If the N\'{e}el vector is in the $x$-$y$ plane with angle $\phi $,\ the
energy spectrum is given by%
\begin{eqnarray}
E^{2} &=&M^{2}\left( \mathbf{k}\right) +\left( \lambda \sin k_{x}+J\cos
^{2}\phi \left( \cos k_{x}-\cos k_{y}\right) \right) ^{2}  \notag \\
&&+\left( \lambda \sin k_{y}+J\sin ^{2}\phi \left( \cos k_{x}-\cos
k_{y}\right) \right) ^{2}.
\end{eqnarray}%
The bulk band gap closes at $J=J_{\text{cr}}$ with 
\begin{equation}
\left\vert \frac{J_{\text{cr}}}{\lambda }\right\vert =\sqrt{\frac{m\left( 2m+%
\sqrt{2}F\right) \sec ^{2}2\phi }{2\left( 4t^{2}-m^{2}\right) }},
\label{Jcr2}
\end{equation}%
and $F\equiv \sqrt{4t^{2}+m^{2}+\left( 4t^{2}-m^{2}\right) \cos 4\phi }$. We
have shown the gap-closing curves in Fig.\ref{FigGap}(b). We later argue
that they are topological phase boundaries when the N\'{e}el vector is
within the $x$-$y$ plane. 
\begin{figure}[t]
\centerline{\includegraphics[width=0.48\textwidth]{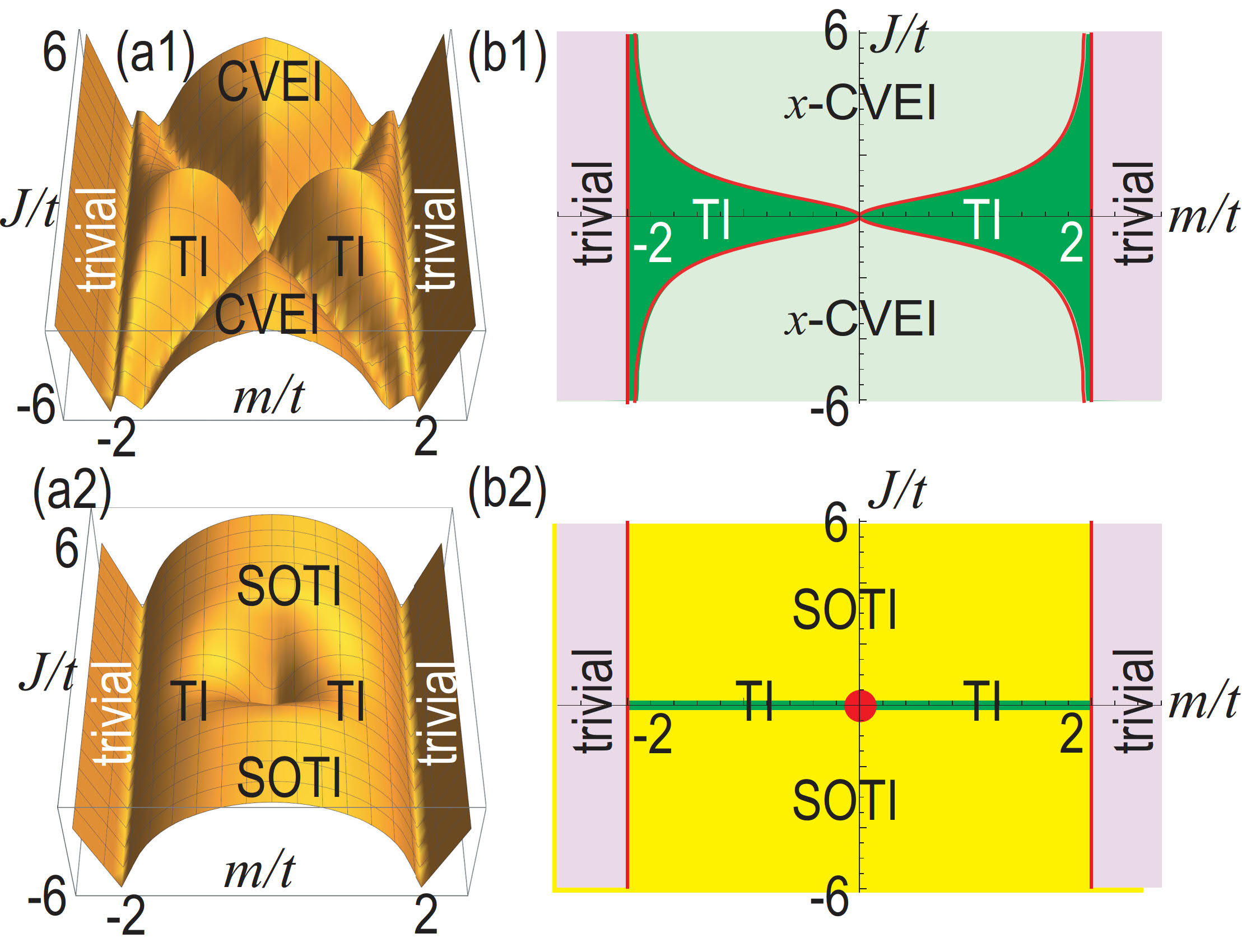}}
\caption{(a1) Bulk band gap in the $m$-$J$ plane, where the N\'{e}el vector
is along the $x$ axis. (b1) Topological phase diagram in the $m$-$J$ plane,
where the N\'{e}el vector is along the $x$ axis. The phase boundaries are
analytically given by Eq.(\protect\ref{Jcr1}). (a2) Bulk band gap in the $m$-%
$J$\ plane, where the N\'{e}el vector is along the $z$\ axis. (b2)
Topological phase diagram in the $m$-$J$\ plane, where the N\'{e}el vector
is along the $z$\ axis. Red disk indicates the gap closing point.}
\label{FigDiagram}
\end{figure}

We show the band-gap structure in the $m$-$J$\ plane when the N\'{e}el
vector is along the $x$-axis in Fig.\ref{FigDiagram}(a1) and the gap-closing
curves in Fig.\ref{FigDiagram}(b1). We later argue that they are topological
phase boundaries. The trivial phase of the BHZ model remains as it is for $%
\left\vert m/\left( 2t\right) \right\vert >1$ and\ $J\neq $0. However, a new
type of trivial insulator emerge for $\left\vert m/\left( 2t\right)
\right\vert <1$\ and $J\neq 0$, as described below.

If the N\'{e}el vector is along the $z$ axis ($\theta =0$), the energy
spectrum is given by%
\begin{equation}
E^{2}=M^{2}\left( \mathbf{k}\right) +\lambda ^{2}\left( \sin ^{2}k_{x}+\sin
^{2}k_{y}\right) +J^{2}\left( \cos k_{x}-\cos k_{y}\right) ^{2}.
\end{equation}%
The band gap is shown in Fig.\ref{FigDiagram}(a2). It implies that the
altermagnetization does not contribute to the gap closing and the gap
closing condition is identical to that of the BHZ model, i.e., $m=\pm 2t$ as
in Fig.\ref{FigDiagram}(b2). Hence, the topological phase boundaries remain
the same as those of the BHZ model. However, the topological phase is
changed from the first-order one to the second-order one, as described below.

\begin{figure}[t]
\centerline{\includegraphics[width=0.48\textwidth]{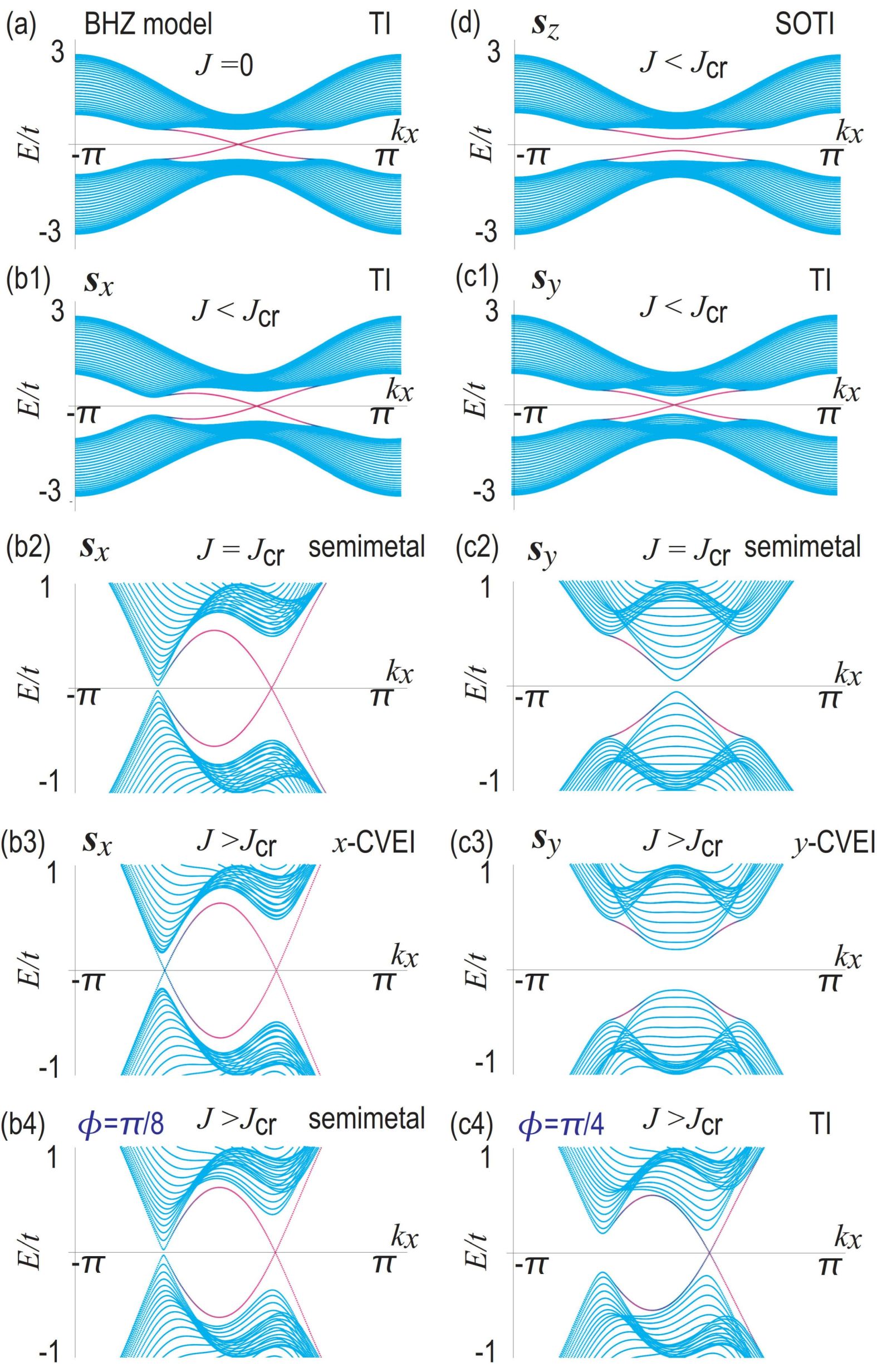}}
\caption{Band structure in ribbon geometry for (a) no altermagnetization,
(b1)$\sim $(b3) the N\'{e}el vector along the $x$ axis, (c1)$\sim $(c3) the N%
\'{e}el vector along the $y$ axis, (b4) the N\'{e}el vector with angle $%
\protect\phi =\protect\pi /8$, (c4) the N\'{e}el vector with angle $\protect%
\phi =\protect\pi /4$, and (d) the N\'{e}el vector along the $z$ axis. The
ribbon is taken along the $x$ axis. The horizontal axis is the momentum $%
k_{x}$. (b1) and (c1) $J=0.2t$. (b2) and (c2) $J=0.5t=J_{\text{cr}}$. (b3),
(c3),\ (b4) and (c4) $J=0.7t$. We have set $m/t=1$, $\protect\lambda /t=0.5$%
. Red color indicates the edge state, while cyan color indicates the bulk
state. The point ($\protect\phi $,$J$) of each figure except (d) is
indicated as a star in the phase diagram Fig.\protect\ref{FigGap}(b). }
\label{FigRibbon}
\end{figure}

\textbf{Ribbon geometry:} We analyze the energy spectrum of the total
Hamiltonian with ribbon geometry to see the topological property. In the
absence of the altermagnetization, $J=0$, there are topological edge states
of the BHZ model (\ref{BHZ}), which cross at the momentum $k_{x}=0$, as in
Fig.\ref{FigRibbon}(a).

If the altermagnetization is along the $x$ axis, $J\mathbf{s}_{x}=\left(
J,0,0\right) $, the crossing point of the topological edge states moves away
from $k_{x}=0$ for $J<J_{\text{cr}}$ as in Fig.\ref{FigRibbon}(b1). At the
critical point $J=J_{\text{cr}}$, the bulk band gap closes and the system is
a semimetal as in Fig.\ref{FigRibbon}(b2). For $J>J_{\text{cr}}$, each edge
state connects either the occupied band or the unoccupied band, as in Fig.%
\ref{FigRibbon}(b3). It indicates the system is trivial. Although the edge
states are nontopological, they are robust against the order of the band gap
because the turning point of the edge state much exceeds the band gap.
Indeed, it requires perturbation much larger than the band gap to remove the
trivial edge states. These nontopological edge states are as robust as
topological edges. They are a reminiscence of the valley-protected edge
states with each edge connecting two valleys either in the occupied band or
in the unoccupied band\cite{Kirch}.

If the altermagnetization is along the $y$ axis, $J\mathbf{s}_{y}=\left(
0,J,0\right) $, the topological edge states are maintained for $J<J_{\text{cr%
}}$ as in Fig.\ref{FigRibbon}(c1). At the critical point $J=J_{\text{cr}}$,
the bulk band gap closes as in Fig.\ref{FigRibbon}(c2). Edge states
disappear for $J>J_{\text{cr}}$ as in Fig.\ref{FigRibbon}(c3).

We show the band structure when the altermagnetization is given by $J\mathbf{%
s}=J\left( \cos \phi ,\sin \phi ,0\right) $\ for the cases $\phi =\pi /8$,\ $%
\pi /4$\ in Fig.\ref{FigRibbon}(b4), (c4), respectively. The bulk band gap
closes in the vicinity of $\phi =\pi /8$ as in Fig.\ref{FigRibbon}(b4) and
Fig.\ref{FigGap}(b). The valley-protected edge states become the topological
edge states as in Fig.\ref{FigRibbon}(c4) and Fig.\ref{FigGap}(b).

If the altermagnetization is along the $z$ axis, $Js_{z}=\left( 0,0,J\right) 
$, the topological edge states anticross and a finite gap emerges in edge
states as in Fig.\ref{FigRibbon}(d). It means that the system is not a
first-order topological insulator. We will soon see that the system is a
second-order topological insulator, which is characterized by the emergence
of corner states in square geometry.

\textbf{Square geometry:} We investigate the energy spectrum of the total
Hamiltonian with square geometry to see the topological property more in
detail.

\begin{figure}[t]
\centerline{\includegraphics[width=0.48\textwidth]{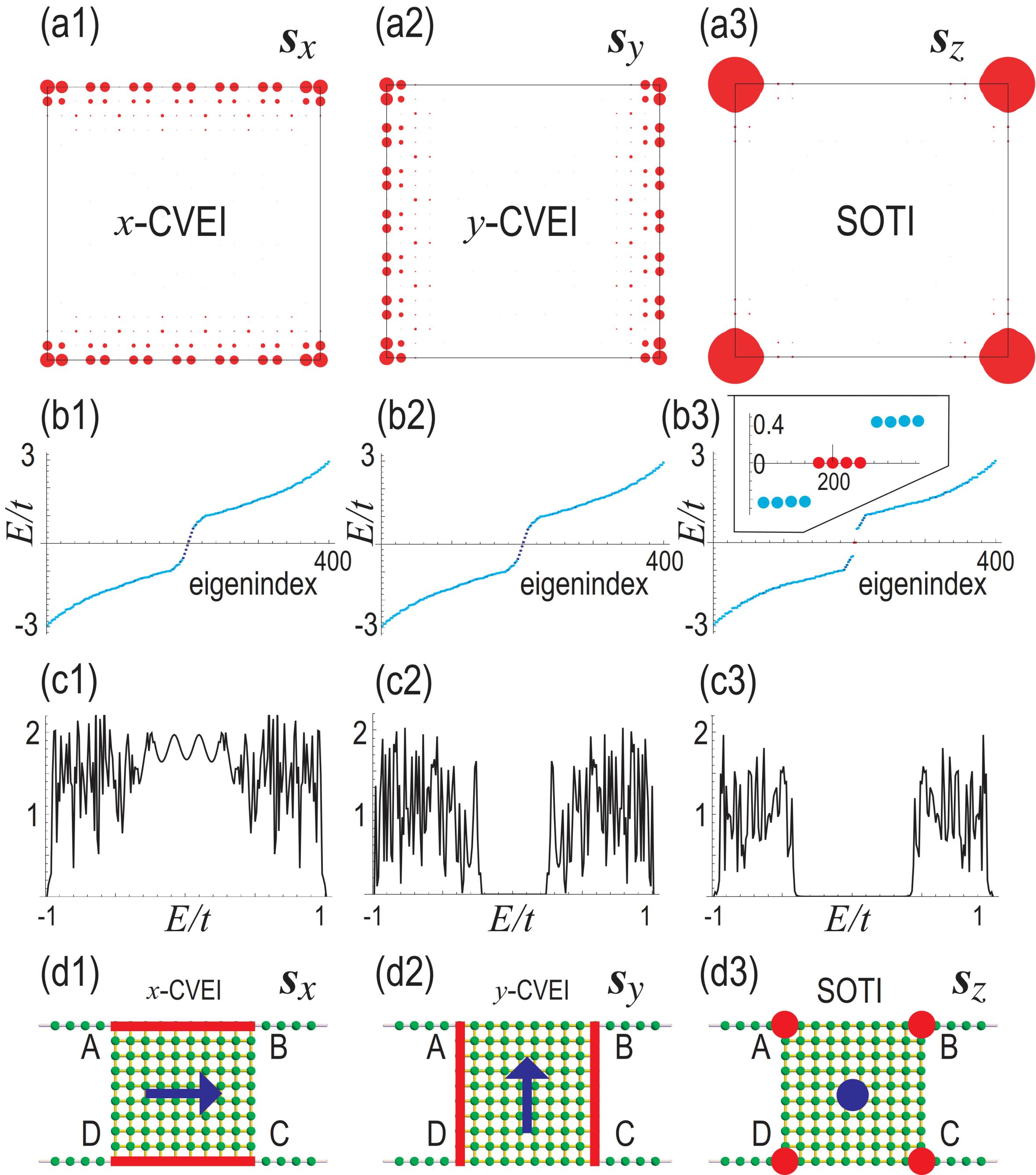}}
\caption{(a1)$\sim $(a3) Spatial distribution of the zero-energy states
marked in red. (b1)$\sim $(b3) Energy spectrum as a function of the
eigenindex of the Hamiltonian. The inset in (b3) shows the presence of the
four corner states at zero energy. (c1)$\sim $(c3) Tunneling conductance as
a function of the energy. (d1)$\sim $(d3) Illustration of a square sample
with four leads. (a1)$\sim $(d1) The N\'{e}el vector is along the $x$ axis.
(a2)$\sim $(d2) The N\'{e}el vector is along the $y$ axis. (a3)$\sim $(d3)
The N\'{e}el vector is along the $z$ axis. We have set $m/t=1$, $\protect%
\lambda /t=0.5$ and $J=0.7t>J_{\text{cr}}$. }
\label{FigBHZ}
\end{figure}

In the case of $J\mathbf{s}_{x}=(J,0,0)$ or $J\mathbf{s}_{y}=\left(
0,J,0\right) $, the edge states emerge along the $x$ axis or the $y$ axis,
as in Fig.\ref{FigBHZ}(a1) or (a2), respectively. Correspondingly, the
energy spectrum is linear as a function of the eigenindex in the vicinity of
the zero energy, as in Fig.\ref{FigBHZ}(b1) or (b2). These edge states are
nontopological as we have argued in ribbon geometry. On the other hand,
there are no edge states along the $y$ axis or the $x$ axis in the case of $J%
\mathbf{s}_{x}=(J,0,0)$ or $J\mathbf{s}_{y}=\left( 0,J,0\right) $ as in Fig.%
\ref{FigBHZ}(a1) or (a2), respectively. It is a reminiscence of edge states
of topological crystalline insulator\cite{EzawaTCI}. The system is the\ $x$%
-axis crystalline valley-edge insulator ($x$-CVEI) in Fig.\ref{FigBHZ}(a1)
and the $y$-axis crystalline valley-edge insulator ($y$-CVEI) in Fig.\ref%
{FigBHZ}(a2).

\begin{figure}[t]
\centerline{\includegraphics[width=0.48\textwidth]{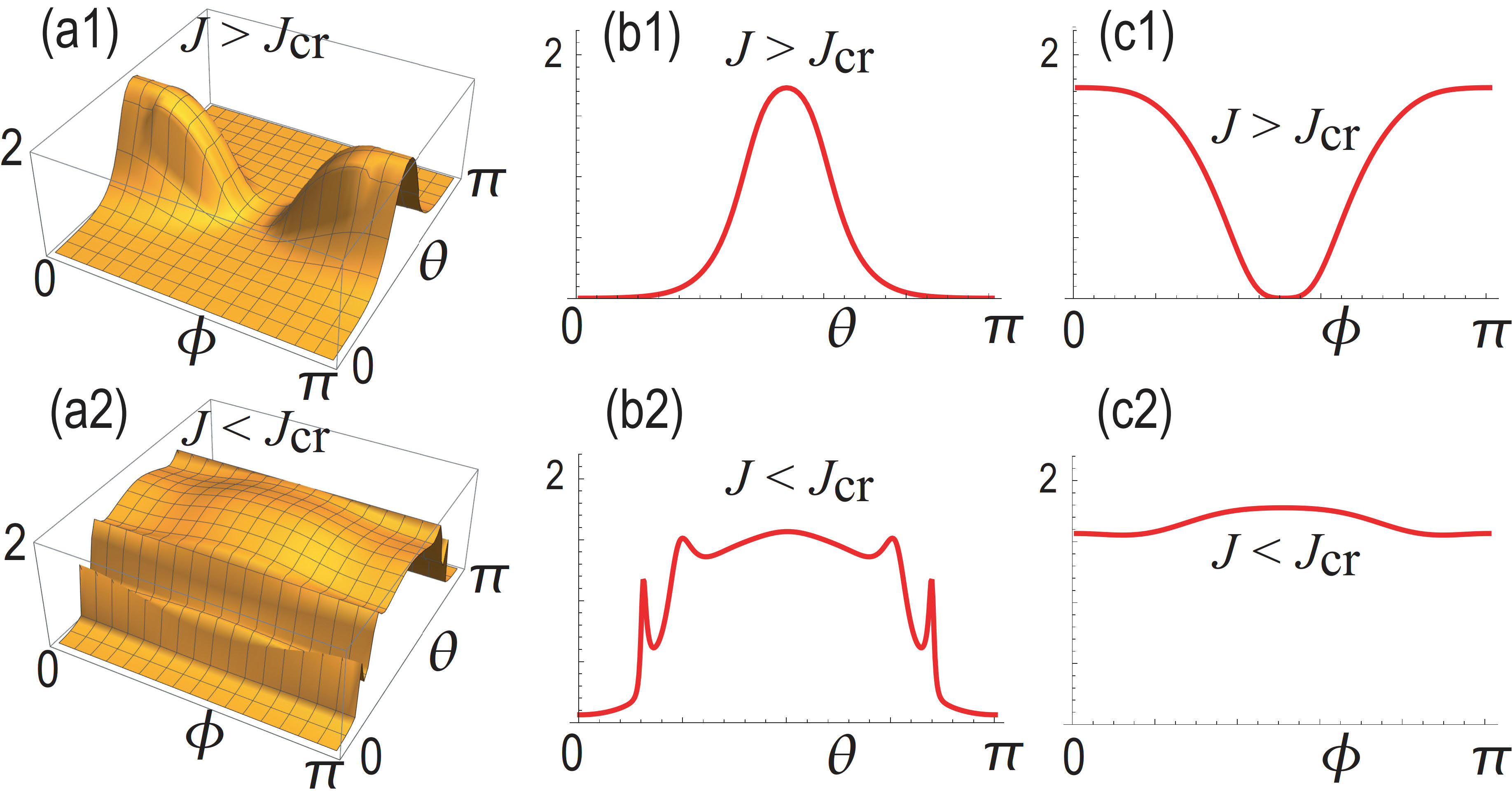}}
\caption{ (a1) and (a2) Tunneling conductance in the $\protect\phi $-$%
\protect\theta $ plane. (b1) and (b2) Tunneling conductance as a function of 
$\protect\theta $ at $\protect\phi =0$. The horizontal axis is $0\leq 
\protect\theta \leq \protect\pi $. (c1) and (c2) Tunneling conductance as a
function of $\protect\phi $ at $\protect\theta =\protect\pi /2$. The
horizontal axis is $0\leq \protect\phi \leq \protect\pi $. (a1)$\sim $(c1) $%
J=0.7t>J_{\text{cr}}$. (a2)$\sim $(c2) $J=0.2t<J_{\text{cr}}$. }
\label{FigConducAngle}
\end{figure}

In the case of $J\mathbf{s}_{z}=\left( 0,0,J\right) $, four topological
corner states emerge as in Fig.\ref{FigBHZ}(a3). Correspondingly, there are
four zero-energy corner states within a finite gap as in Fig.\ref{FigBHZ}%
(b3). As a result, the system is a second-order topological insulator.

\textbf{Conductance:} We show that the N\'{e}el vector is detectable by
measuring the conductance in the setup shown in Fig.\ref{FigBHZ}(d1)$\sim $%
(d3), where four leads are attached to the corners of the square sample. We
assume that the leads are\ single-atomic chains with semi-infinite length.

The conductance between two leads is calculated based on the Landauer
formalism\cite{Datta,Rojas,Nikolic,Li,EzawaSwitch}. The conductance $\sigma
(E)$ at energy $E$ is calculated as\cite{Datta} 
\begin{equation}
\sigma (E)=(e^{2}/h)\text{Tr}[\Gamma _{\text{L}}(E)G_{\text{D}}^{\dag
}(E)\Gamma _{\text{R}}(E)G_{\text{D}}(E)],  \label{G}
\end{equation}%
where $\Gamma _{\text{R(L)}}(E)=i[\Sigma _{\text{R(L)}}(E)-\Sigma _{\text{%
R(L)}}^{\dag }(E)]$ is the line width with the self-energies $\Sigma _{\text{%
L}}(E)$ and $\Sigma _{\text{R}}(E)$ for the left and right leads, and $G_{%
\text{D}}(E)=[E-H-\Sigma _{\text{L}}(E)-\Sigma _{\text{R}}(E)]^{-1}$ is the
Green function with the Hamiltonian $H$ for the sample. The self energy of a
single-atomic semi-infinite chain is analytically obtained\cite{Datta} as $%
\Sigma _{\text{L}}(E)=\Sigma _{\text{R}}(E)=E-i\sqrt{\left\vert
t^{2}-E^{2}\right\vert }$.

First, we study the case $J>J_{\text{cr}}$. We show the conductance as a
function of the energy $E$ in Fig.\ref{FigBHZ}(c1), (c2) and (c3). We have
calculated the conductance between the leads A and B. The conductance is
nonzero at $E=0$ when the N\'{e}el vector is along the $x$ axis as in Fig.%
\ref{FigBHZ}(c1). It is understood that the current flows through the edge
state along the $x$ axis as in Fig.\ref{FigBHZ}(d1). On the other hand, it
is zero at $E=0$ when the N\'{e}el vector is along the $y$ axis as in Fig.%
\ref{FigBHZ}(c2). It\ is because there is no edge state to carry the current
along the $x$\ axis as in Fig.\ref{FigBHZ}(d2). The conductance is also zero
at $E=0$\ when the N\'{e}el vector is along the $z$\ axis as in Fig.\ref%
{FigBHZ}(c3). It is because there is no edge state although corner states
exist as in Fig.\ref{FigBHZ}(d3). It seems to be difficult to differentiate
the cases where the N\'{e}el vector is along the $y$\ axis or the $z$\ axis
by measuring the conductance because the tunneling conductance is zero for
both cases. However, it is differentiated by mearing the tunneling
conductance between the leads A and D, where the conductance is nonzero if
the N\'{e}el vector is along the $y$\ axis but zero if the N\'{e}el vector
is along the $z$\ axis.

The conductance in the $\phi $-$\theta $ plane is shown in Fig.\ref%
{FigConducAngle}(a1). The conductance as a function of $\theta $ at $\phi =0$
is shown in Fig.\ref{FigConducAngle}(b1). The conductance takes maximum
value at $\theta =\pi /2$ and takes minimum value at $\theta =0$. The
conductance as a function of $\phi $ at $\theta =\pi /2$ is shown in Fig.\ref%
{FigConducAngle}(c1). The conductance takes the maximum value at $\phi =0$
and the minimum value at $\phi =\pi /2$. Hence, we can detect the N\'{e}el
vector by measuring the conductance.

Next, we study the case $J<J_{\text{cr}}$. We show the conductance in the $%
\phi $-$\theta $ plane in Fig.\ref{FigConducAngle}(a2). The dependence on
the angle $\phi $ is tiny comparing with the case $J>J_{\text{cr}}$, as in
Fig.\ref{FigConducAngle}(c2). It is because the system is a first-order
topological insulator, where the topological edge state surround the sample.
The conductance is zero for $\theta =0$ and $\pi $ as in the case of the
system $J>J_{\text{cr}}$, as in Fig.\ref{FigConducAngle}(b2). It is because
there are topological corner states but there are no edge states.

\begin{figure}[t]
\centerline{\includegraphics[width=0.48\textwidth]{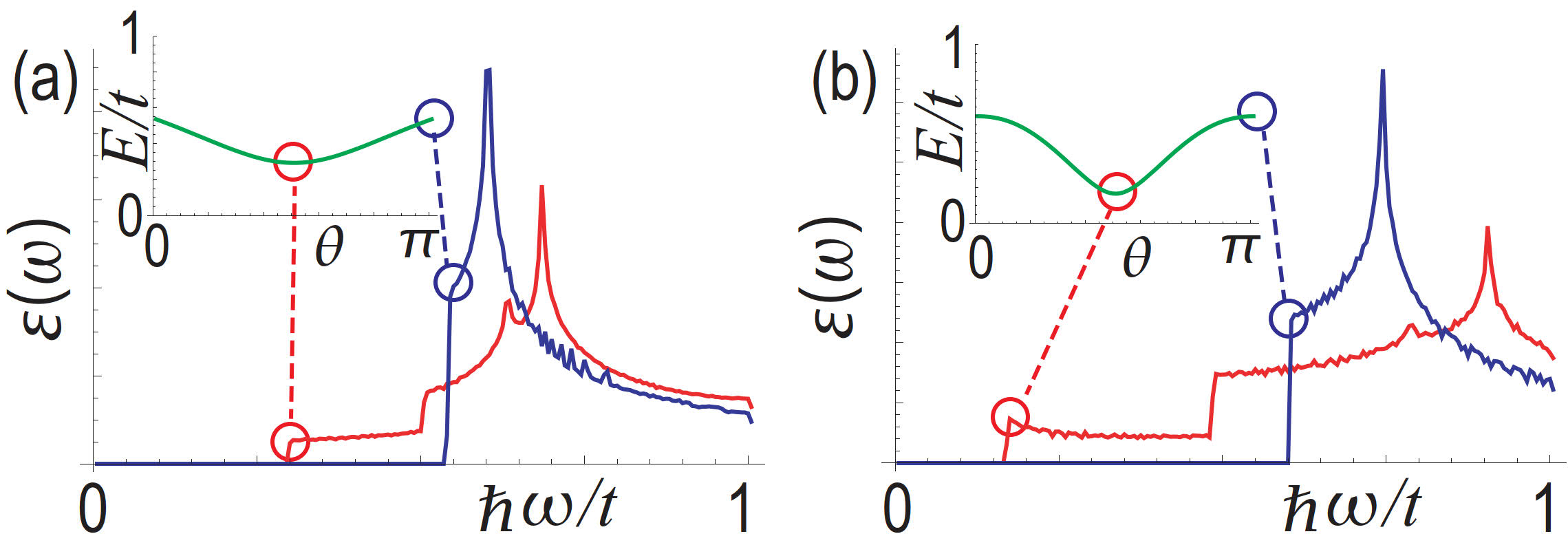}}
\caption{Optical absorption spectra. Red curves indicate the spectra with $%
\protect\theta =\protect\pi /2$, while blue curves indicate that with $%
\protect\theta =0$. We have set $\protect\phi =0$. The horizontal axis is
the photon energy $\hbar \protect\omega $. Insets show the bulk band gap as
a function of $\protect\theta $ at $\protect\phi =0$. (a) $J=0.2t<J_{\text{cr%
}}$ and (b) $J=0.7t>J_{\text{cr}}$. The red and blue circles correspond to
the optical absorption colored in red and blue.}
\label{FigOpt}
\end{figure}

\textbf{Optical absorption:} We show that the band gap is observed by
examining the optical absorption spectrum. We analyze the optical inter-band
transition from the state $|u_{\text{v}}(\mathbf{k})\rangle $\ in the
valence band to the state $|u_{\text{c}}(\mathbf{k})\rangle $\ in the
conduction band. We apply a circularly polarized light, where the
electromagnetic potential is given by $A(t)=(A_{x}\sin \omega t,A_{y}\cos
\omega t)$.

The optical absorption is calculated based on the Kubo formula\cite%
{YaoOpt,XiaoOpt,LiOpt,EzawaOpt,EzawaTCI},%
\begin{align}
\varepsilon \left( \omega \right) =& \frac{\pi e^{2}}{\varepsilon
_{0}m_{e}^{2}\omega ^{2}}\sum_{i}\int_{BZ}\frac{d\mathbf{k}}{\left( 2\pi
\right) ^{2}}f(\mathbf{k})\left\vert P_{j}(\mathbf{k})\right\vert ^{2} 
\notag \\
& \times \delta \left[ E_{\text{c}}(\mathbf{k})-E_{\text{v}}(\mathbf{k}%
)-\hbar \omega \right] ,  \label{absorp}
\end{align}%
where $P_{j}(\mathbf{k})$ is the optical matrix element $P_{j}(\mathbf{k}%
)\equiv m_{0}\left\langle u_{\text{c}}(\mathbf{k})\right\vert \frac{\partial
H}{\partial k_{j}}\left\vert u_{\text{v}}(\mathbf{k})\right\rangle $, $E_{%
\text{c}}(\mathbf{k})$\ and $E_{\text{v}}(\mathbf{k})$\ are the energies of
the conduction and valence bands, and $f(\mathbf{k})$\ is the Fermi
distribution function.

We show the optical absorption spectra at $\theta =0$ and\ $\pi /2$\ in Fig.%
\ref{FigOpt}. There is a gap in the spectrum, implying the sudden occurrence
of the optical absorption at the moment that the photon energy becomes the
same as the band-gap energy. Hence, the band gap is measured by optical
absorption experiments. 

In conclusion, it is possible to determine the N\'{e}el vector by measuring
the tunneling conductance and the optical absorption.

The author is very much grateful to S. Seki for helpful discussions on the
subject. This work is supported by CREST, JST (Grants No. JPMJCR20T2) and
Grants-in-Aid for Scientific Research from MEXT KAKENHI (Grant No.
23H00171). 

\end{document}